\documentclass[runningheads]{llncs}
\usepackage[T1]{fontenc}
\usepackage{graphicx}
\usepackage{hyperref}
\usepackage{xcolor}
\usepackage{multirow}
\usepackage{amsmath}
\usepackage{booktabs}

\usepackage[normalem]{ulem}
\useunder{\uline}{\ul}{}
\usepackage[T1]{fontenc}
\usepackage{lmodern}

\newcommand{\lambdavar}{\ensuremath{\lambda_{\textit{var}}}}
\newcommand{\Lsiindep}{\ensuremath{\mathcal{L}_{\textit{si}}^{\textit{indep}}}}
\newcommand{\Lsidep}{\ensuremath{\mathcal{L}_{si}^{\textit{dep}}}}
\newcommand{\Lcasiindep}{\ensuremath{\mathcal{L}_{\textit{casi}}^{\textit{indep}}}}
\newcommand{\Lcasidep}{\ensuremath{\mathcal{L}_{\textit{casi}}^{\textit{dep}}}}

\begin{document}
\newcommand{\methodname}{3DDX}
\title{\methodname: Bone Surface Reconstruction\\from a Single Standard-Geometry Radiograph\\via Dual-Face Depth Estimation}
\titlerunning{\methodname: Bone surface reconstruction from a single radiograph}

\author{
Yi Gu\inst{1} \and
Yoshito Otake\inst{1} \and
Keisuke Uemura\inst{2} \and
Masaki Takao\inst{3} \and
Mazen Soufi\inst{1} \and\\
Seiji Okada\inst{4} \and
Nobuhiko Sugano\inst{2} \and
Hugues Talbot\inst{5} \and
Yoshinobu Sato\inst{1}
}

% index{Gu, Yi}
% index{Otake, Yoshito}
% index{Uemura, Keisuke}
% index{Takao, Masaki}
% index{Soufi, Mazen}
% index{Okada, Seiji}
% index{Sugano, Nobuhiko}
% index{Talbot, Hugues}
% index{Sato, Yoshinobu}

%
\authorrunning{Y. Gu et al.}
% \institute{
% Graduate School of Science and Technology, \\
% Nara Institute of Science and Technology, Japan\\
% \email{gu.yi.gu4@naist.ac.jp} \and
% Department of Orthopeadic Medical Engineering, \\
% Osaka University Graduate School of Medicine, Japan \and
% Department of Bone and Joint Surgery, \\
% Ehime University Graduate School of Medicine, Japan \and
% Department of Orthopaedics, \\
% Osaka University Graduate School of Medicine, Japan \and
% CentraleSupélec, Université Paris-Saclay, France
% }

% \institute{
% Nara Institute of Science and Technology, Japan\\
% \email{gu.yi.gu4@naist.ac.jp} \and
% Osaka University Graduate School of Medicine, Japan \and
% Ehime University Graduate School of Medicine, Japan \and
% CentraleSupélec, Université Paris-Saclay, France
% }

\institute{
Division of Information Science, Graduate School of Science and Technology, \\
Nara Institute of Science and Technology, Japan\\
\email{gu.yi.gu4@naist.ac.jp, \{otake,yoshi\}@is.naist.jp} \and
Department of Orthopeadic Medical Engineering, \\
Osaka University Graduate School of Medicine, Japan \and
Department of Bone and Joint Surgery, \\
Ehime University Graduate School of Medicine, Japan \and
Department of Orthopaedics,
Osaka University Graduate School of Medicine, Japan \and
CentraleSupélec, Université Paris-Saclay, France
}

\maketitle              % typeset the header of the contribution
\begin{abstract}
Radiography is widely used in orthopedics for its affordability and low radiation exposure. 3D reconstruction from a single radiograph, so-called 2D-3D reconstruction, offers the possibility of various clinical applications, but achieving clinically viable accuracy and computational efficiency is still an unsolved challenge. Unlike other areas in computer vision, X-ray imaging's unique properties, such as ray penetration and fixed geometry, have not been fully exploited. We propose a novel approach that simultaneously learns multiple depth maps (front- and back-surface of multiple bones) derived from the X-ray image to computed tomography registration. The proposed method not only leverages the fixed geometry characteristic of X-ray imaging but also enhances the precision of the reconstruction of the whole surface. Our study involved 600 CT and 2651 X-ray images (4 to 5 posed X-ray images per patient), demonstrating our method's superiority over traditional approaches with a surface reconstruction error reduction from 4.78 mm to 1.96 mm. This significant accuracy improvement and enhanced computational efficiency suggest our approach's potential for clinical application.

\keywords{Monocular depth estimation \and X-ray radiography \and deep learning \and inverse problems}
\end{abstract}
\section{Introduction}
% Depth map estimation in computer vision has many applications and explored extensively, but the scale-invariance constraint has been usually employed to robustify the estimation. The scale has been predicted from other sources such as....

% Algorithms for 2D-3D reconstruction from a single or multiple images either used fitting of statistical model to the 2D image or trained the mapping between 2D image and 3D volume, but they were low accuracy (smoothed surface close to the mean shape) or low resolution due to the limitation of computational resources.

% In this paper, we propose 2D-3D reconstruction algorithm that predicts a 3D shape of the examined body parts from a single diagnostic radiograph obtained in a standardized geometry.

Achieving monocular or 2D-3D reconstruction is a long-standing challenge in computer vision and medical engineering.
Practical 3D reconstruction from radiographs has recently been a hot topic considering the significance of clinical applications.
% Usually, multiple radiographs are necessary to perform 3D reconstruction \cite{baka_2d_3d_2011,youn_iterative_2017,balestra_articulated_2014,chenes_revisiting_2021,almeida_three_dimensional_2021,cafaro_x2vision_2023,kasten_end2end_2020}, with chest tomosynthesis as a major application~\cite{dobbins2009chest}.
Usually, multiple radiographs are necessary to perform 3D reconstruction \cite{baka_2d_3d_2011,youn_iterative_2017,balestra_articulated_2014,chenes_revisiting_2021,almeida_three_dimensional_2021,cafaro_x2vision_2023,kasten_end2end_2020,dobbins2009chest}.
Only a few works have tried to achieve 3D reconstruction using single X-ray images \cite{shiode_2d3d_2021,ha_2d_3d_2024,jiang_reconstruction_2021,tan_xctnet_2022}.
However, existing works suffer from low reconstruction quality, accuracy, and resolution as well as high computational cost, which significantly limit clinical applications.
On the other hand, monocular depth estimation from a single camera image~\cite{eigen_depth_2014}, which offers impressive 3D reconstruction, has been extensively studied and widely applied, becoming an essential part of many vision models~\cite{liu_multi_modal_2023,xiang_3d_aware_2023,wang_tracking_2023}.
Nevertheless, the relation between depth map and X-ray image has barely been explored, especially for the topic of 2D-3D reconstruction.

In this paper, we shed light on a new path to the 3D reconstruction from a single X-ray image, using depth estimation.
Realizing the unique properties of penetrating rays in X-ray imaging, we propose simultaneous 3D dual-face (front and back) depth estimation from a single X-ray image (\methodname), for 3D reconstruction.
In the classic monocular depth estimation problem, the relative depth estimation (RDE) \cite{ranftl_towards_2022,yang_depth_2024}, which only cares about relative depth, and metric depth estimation (MDE), which estimates absolute physical-unit depth \cite{eigen_depth_2014,lee_big_2021,bhat_adabins_2021,avidan_localbins_2022,bhat_zoedepth_2023} are two major task categories.
We focus on MDE for meaningful clinical application with physical units.
However, conventional losses were designed for estimating single depth maps, where we try to estimate multiple depth maps from a single input.
To tackle that, we propose a generalization of the loss functions to multi-depth-map supervision.
Furthermore, we take advantage of a fixed imaging geometry, namely the relative position of the X-ray source with respect to the detector, by realizing that the diagnostic radiography is standardized \cite{clohisy_systematic_2008}. 
% For example, Clohisy et al. \cite{clohisy_systematic_2008} recommended a source-to-detector distance of 1200 mm for hip examinations, which is the standard setup in most clinical examinations.
To the best of our knowledge, we are the first to achieve 3D bone reconstruction from a single X-ray image acquired in a clinical setup using depth estimation.\\
% In addition to the accuracy, our method basically works with 2D images for 3D reconstruction, which is highly efficient at both training and inference stages.
% We believe the proposed method entails potential for wild clinical applications such as 3D bone angle estimation and surgical planing. 
% \subsubsection{Contribution}
Contribution: We propose a method (\methodname) for the reconstruction of 3D bone surfaces with absolute scaling and large field-of-view, while retaining high-resolution details from single X-ray images acquired in a clinically standardized geometric setup. Our contribution is three-fold: 1) proposal of a dual-face depth estimation from a single X-ray image by exploiting information from the penetrating X-ray, 2) proposal of a new loss function in a depth map estimation network allowing the scale-specific training under a specific geometric constraint, 3) extensive evaluation using a large-scale hip X-ray image database (600 patients, 2651 X-ray images) paired with CT image through 2D-3D registration. Our code is available at \url{https://github.com/Kayaba-Akihiko/3DDX}.

\begin{figure}
\includegraphics[width=\textwidth]{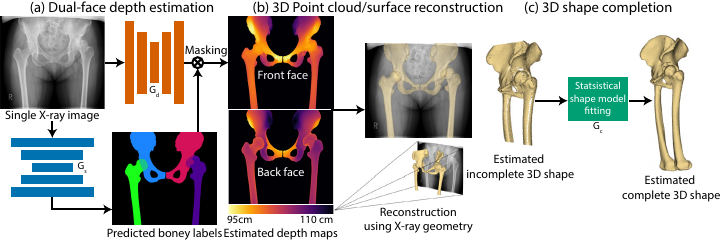}
\caption{Overview of the proposed method. (a) The dual-face depth estimation using depth maps estimation model $G_{d}$ and a bone segmentation model $G_{s}$ to mask the invalid region. (b) 3D surface reconstruction from the estimated depth maps using X-ray geometry to produce initial 3D estimation. (c) 3D shape completion using bone statistical shape model fitting $G_{c}$.} \label{fig:method_overview}
\end{figure}

\section{Method}
Fig. \ref{fig:method_overview} shows an overview of the proposed method. We build a novel framework for estimating the complete 3D shape of the femur and pelvis (including unseen regions) from a single X-ray image. To this end, we propose to estimate front- and back-face depth maps for each target object (e.g., a hemipelvis) for the 3D reconstruction from a plain X-ray image.
A depth maps estimation model $G_{d}$ is trained to estimate all the depth maps.
We propose a simple yet effective loss function to improve the depth estimation performance by leveraging the standardized geometry information in X-ray imaging.
We also train a segmentation model to generate the masks of target objects from an X-ray image, masking the invalid region (i.e., the non-target region) on the estimated depth maps for bone reconstruction.
Using the given X-ray image geometry, the point cloud (PCD) of bone is constructed from the estimated depth maps.
We perform 3D shape completion with the statistical shape model (SSM) fitting $G_{c}$ to further validate the superiority of using dual-face depth.

\subsection{Depth maps estimation}
We revisited a popular MDE loss, scale-invariant (SI) loss \cite{eigen_depth_2014,lee_big_2021,bhat_adabins_2021} that preserves learning the global scale and shift for estimating a depth map, defined as
% The scale-invariant loss \cite{lee_big_2021,bhat_adabins_2021} is defined as
\begin{equation}
\mathcal{L}_{si}=\alpha\sqrt{D(g)} = \alpha\sqrt{\frac{1}{T}\sum_i{g^2_i}-\frac{\lambdavar}{T^2}(\sum_i{g_i})^2},
\end{equation}
where $g_i=\log{\hat{y}_i}-\log{y_i}$ is the error logarithm between the $i-$th predicted depth $\hat{y}_i$ and
ground truth depth $y_i$, assuming $T$ valid pixels.
Following \cite{lee_big_2021,bhat_adabins_2021}, all experiments set the $\alpha$ and $\lambdavar$ to 10 and 0.85, respectively.
% For the MDE tasks, depth bin techniques \cite{avidan_localbins_2022,bhat_zoedepth_2023,bhat_adabins_2021}, which transform depth regression task into a classification task, are often used.
% However, those techniques require additional learnable modules, increasing computational cost.
% For the MDE tasks, the costly depth bin techniques \cite{avidan_localbins_2022,bhat_zoedepth_2023,bhat_adabins_2021} are often used.
This work focuses on improving the SI loss for multiple depth map supervision.
Generalization of the SI loss to multiple depth maps leads to multiple functions, considering the inter-depth-map pixels relations since the loss considers pixel-to-pixel relations by minimizing the error variance.
In the following subsections, we proposed the SI loss generalizations and improvement.
We will discuss the performance difference in the results section Sec. \ref{sec:exp}.
\subsubsection{Generalization to multiple depth maps}
Eq. \eqref{eq:si_indep} and \eqref{eq:si_dep} are two straightforward ways to generalize the SI loss $\mathcal{L}_{si}$.
In particular, \eqref{eq:si_indep} is a simple averaging of the SI losses of all depth maps, where $g^j$ is the log error between the $j-$th ground truth and estimated depth maps in $N$ pairs. 
In this way, the inter-depth-map pixels are independent of each other as the error variance is calculated separately.
For considering inter-depth-map pixels relation,  Eq. \eqref{eq:si_dep} is presented, where the $T^j$ is the number of valid pixels in $j-$th depth map. 
% The expectation might be that $\mathcal{L}_{si}^{dep}$ is better; however, our experiments show that $\mathcal{L}_{si}^{indep}$ outperforms $\mathcal{L}_{si}^{dep}$ and is more stable. We will show the detailed comparison in the Result section. The possible reason for the inferiority of the $\mathcal{L}_{si}^{dep}$ might be that the different objects' depth (pelvis and femur in our case) are statistically different, resulting in different variance level.
\begin{equation}
    \Lsiindep=\frac{\alpha}{N}\sum_j\sqrt{D(g^j)}
\label{eq:si_indep}
\end{equation}
\begin{equation}
    \Lsidep=\alpha\sqrt{M(g)}=\alpha\sqrt{\frac{1}{\sum_jT^j}\sum_j\sum_i(g_i^j)^2-\frac{\lambdavar}{(\sum_jT^j)^2}(\sum_j\sum_ig_i^j)^2}
\label{eq:si_dep}
\end{equation}

\subsubsection{Center-aligned scale-invariant loss}
The vanilla SI error supervises both scale and shift, which is a general need but not in our case.
To leverage the fixed imaging geometry information, we propose the center-aligned SI loss (CASI), which supervises only the scale while allowing depth shifting by center alignment.
A popular way to align the center is centralizing the prediction and ground truth to the depth origin.
However, the scale-invariant log error only allows positive depth.
Consequently, we propose to align the estimated depth center to the ground truth center using \eqref{eq:casi_error}, which is equivalent to performing \textit{rigid registration} on the depth, where the $t(\cdot)$ calculates the mean of given valid pixels. 
The $(\cdot)^+$ and $\varepsilon$ are the ReLU function and a numerical safeguard, respectively.
The proposed independent and dependent CASI losses were then defined as $\Lcasiindep=\frac{\alpha}{N}\sum_j\sqrt{D(h^j)}$ and $\Lcasidep=\alpha\sqrt{M(h)}$, respectively.
Thus, the proposed CASI loss does not introduce new tuning parameters, which lowers the hyperparameter search burden.
\begin{equation}
    h_i^j=\log{\left(\left(\hat{y}_i^j+t(y)-t(\hat{y})\right)^++\varepsilon\right)}-\log{(y_i^j+\varepsilon)}
    \label{eq:casi_error}
\end{equation}

\subsubsection{Segmentation of depth maps}
We train a segmentation model $G_{s}$ to generate the bone masks for removing the background region in 3D bone surface reconstruction step. We use the Dice semimetric losses \cite{wang_dice_2023} with Cross-Entropy loss and label smoothing \cite{muller_when_2019} for training.
Segmentation is considered a pixel-wise multi-class classification, allowing label overlay (e.g., in the hip joint region).

\subsection{Surface reconstruction and 3D shape completion}
We compare the 3D shape completion performance between the single-face-depth-map-reconstructed 3D shape (the conventional method) and the dual-face-depth-map-reconstructed 3D shape (our proposal).
The object surfaces are reconstructed from estimated depth maps with the predicted bone labels, using fixed imaging geometry.
We perform SSM fitting~\cite{baka_2d_3d_2011,whitmarsh_reconstructing_2011} for 3D shape completion.
We build an SSM for each object we target.
The GBCPD algorithm is used \cite{hirose_geodesic_based_2023} for both rigid and non-rigid registration for constructing point-to-point correspondence.
During the inference, the statistical shapes are fitted to the incomplete shape to estimate the complete shape.
The cost function is defined as
\begin{equation}
    \mathcal{L}_{\textit{ssm}}(\theta)=\text{dist}(\text{clip}(\hat{s}(\theta), s),s)+\frac{\lambda_{l2}}{N_\theta}\sum_i\theta_i^2,
\end{equation}
where $\theta$ is the $N_\theta$-D vector for the optimization for fitting and the second term is a $\lambda_{l2}$-weighted $l2$ regularization.
$\text{clip}(\hat{s}(\theta),s)$ clips the estimated shape $\hat{s}(\theta)$ to as the same field-of-view as the fitting target shape $s$.
The function $\text{dist}(\cdot)$ measures the bi-directional shape distance if the fitting target is built from the proposed dual-face depth maps; otherwise, it measures the directional shape distance from the target to the shape model.
The L-BFGS algorithm \cite{liu_limited_1989} was used to search the optimal $\theta$. The $\lambda_{l2}$ was set to 0.01

{
\DeclareFontSeriesDefault[rm]{bf}{sb}   % semi-bold
%\DeclareFontSeriesDefault[rm]{bf}{x}   % no-bold
%\DeclareFontSeriesDefault[rm]{bf}{bx}   % extended-bold

\begin{table}[]
\caption{
Evaluation results of point cloud reconstruction with and without shape completion.
For shape completion, the healthy and diseased bones are reported separately.
{\scriptsize\textit{256}}, {\scriptsize\textit{512}}, and {\scriptsize\textit{1024}} refer to the X-ray resolutions.
\textreferencemark{ }denotes 3D reconstruction using single-face depth maps.
\textdagger{ }denotes using pretraining.
The mean(std.) of the metrics are reported.
ASSD, HD95, EMD, are reported in mm unit; $\mathrm{CD}_{l2}$ is in $\mathrm{mm^2}$ unit.
}
\label{tb:eval_3d}
\scriptsize
\centering
\begin{tabular}{lcccccccc}
\toprule
% \hline
\multicolumn{9}{c}{\normalsize Point cloud evaluation {\scriptsize $\downarrow$($\downarrow$)}}                                                                                                                                                                             \\ \hline
\multicolumn{1}{l|}{\multirow{2}{*}{Method}}                                                   & \multicolumn{4}{c|}{Pelvis}                                       & \multicolumn{4}{c}{Femur}                         \\
\multicolumn{1}{l|}{}                                                                          & ASSD       & HD95       & EMD        & \multicolumn{1}{c|}{CD-l2} & ASSD       & HD95       & EMD        & CD$_{l2}$      \\ \hline
{\tiny\textit{256}} $\Lsiindep$\textreferencemark  & 4.78(0.85) & 18.0(2.13) & 8.55(1.15) & 115(30.8)                  & 5.54(1.52) & 21.1(2.63) & 9.36(2.09) & 152(68.9)  \\
{\tiny\textit{256}} $\Lsiindep$                    & 2.11(0.77) & 5.82(2.08) & 3.14(2.19) & 21.2(21.5)                 & 2.28(1.60) & 5.75(4.10) & 3.13(2.39) & 25.6(61.9) \\
{\tiny\textit{256}} $\Lcasidep$                    & 1.96(0.77) & 5.38(2.09) & 2.93(1.18) & 18.9(21.2)                 & 2.20(1.66) & 5.60(4.39) & 3.03(2.49) & 25.4(68.7) \\
{\tiny\textit{256}} $\Lcasiindep$                  & \textbf{1.95}(0.78) & \textbf{5.36}(2.10) & \textbf{2.92}(1.18) & \textbf{18.8}(21.3)                 & \textbf{2.15}(1.66) & \textbf{5.49}(4.38) & \textbf{2.97}(2.50) & \textbf{24.7}(68.5) \\ \hline
{\tiny\textit{256}} $\Lcasiindep$\textdagger       & 1.93(0.77) & 5.30(2.11) & 2.88(1.17) & 18.5(21.2)                 & 2.12(1.66) & 5.42(4.42) & 2.93(2.50) & 24.4(77.3) \\
{\tiny\textit{512}} $\Lcasiindep$\textdagger       & 1.80(0.76) & 4.93(2.07) & 2.73(1.16) & 16.9(20.7)                 & 1.99(1.56) & 5.14(4.17) & 2.76(2.39) & 21.8(60.0) \\
{\tiny\textit{1024}} $\Lcasiindep$\textdagger      & \textbf{1.76}(0.75) & \textbf{4.82}(2.02) & \textbf{2.69}(1.13) & \textbf{16.3}(19.3)                 & \textbf{1.95}(1.55) & \textbf{5.07}(4.19) & \textbf{2.71}(2.34) & \textbf{21.2}(61.3) \\ 
\toprule 
% \hline
\multicolumn{9}{c}{\normalsize 3D completion evaluation {\scriptsize $\downarrow$($\downarrow$)}}                                                                                                                                                                           \\ \hline
\multicolumn{1}{l|}{\multirow{2}{*}{\begin{tabular}[c]{@{}l@{}}Fitting\\ target\end{tabular}}} & \multicolumn{4}{c|}{Healthy pelvis}                               & \multicolumn{4}{c}{Healthy femur}                 \\
\multicolumn{1}{l|}{}                                                                          & ASSD       & HD95       & EMD        & \multicolumn{1}{c|}{CDl2}  & ASSD       & HD95       & EMD        & CD${_l2}$       \\ \hline
{\tiny\textit{256}} $\Lsiindep$\textreferencemark & 2.34(0.69) & 5.95(2.14) & 3.13(0.95) & 19.7(17.6)                 & 3.78(2.75) & 9.21(6.87) & 4.88(3.56) & 72.3(148)  \\
{\tiny\textit{256}} $\Lcasiindep$                 & 1.95(0.61) & 5.06(1.99) & 2.73(0.86) & 13.9(15.2)                 & 2.19(1.16) & 5.40(2.86) & 2.93(1.64) & 19.3(35.3) \\
{\tiny\textit{1024}}$\Lcasiindep$\textdagger      & \textbf{1.91}(0.60) & \textbf{4.91}(1.98) & \textbf{2.66}(0.85) & \textbf{13.1}(15.1)                 & \textbf{2.11}(1.15) & \textbf{5.22}(2.83) & \textbf{2.85}(1.60) & \textbf{18.2}(33.7) \\ \hline
\multicolumn{1}{c|}{}                                                                          & \multicolumn{4}{c|}{Diseased pelvis}                              & \multicolumn{4}{c}{Affected femur}                \\ \hline
{\tiny\textit{256}} $\Lsiindep$\textreferencemark & 2.55(0.88) & 6.84(2.85) & 3.50(1.27) & 24.8(23.7)                 & 4.56(3.05) & 11.5(7.71) & 6.11(4.22) & 101(151)   \\
{\tiny\textit{256}} $\Lcasiindep$                 & 2.15(0.80) & 5.80(2.71) & 3.03(1.18) & 17.8(21.7)                 & 2.62(1.68) & 6.67(4.48) & 3.54(2.45) & 31.2(69.5) \\
{\tiny\textit{1024}}$\Lcasiindep$\textdagger      & \textbf{2.05}(0.93) & \textbf{5.63}(2.80) & \textbf{2.95}(1.33) & \textbf{17.5}(37.8)                 & \textbf{2.51}(1.57) & \textbf{6.40}(4.19) & \textbf{3.40}(2.26) & \textbf{28.1}(61.2) \\ \hline
\end{tabular}
\end{table}
}

\section{Experiments and Results}
\label{sec:exp}
We collected 2651 X-ray images (600 patients) paired with their respective CT images.
CT bone segmentation \cite{hiasa_automated_2020} and X-ray 2D-3D registration \cite{otake_intraoperative_2012} were performed to produce ground truth bone 3D shapes and depth maps.
Ethical approval was obtained from the Institutional Review Boards at Osaka University and Nara Institute of Science and Technology (approval numbers 15056-3 and 2019-M-6, respectively).
We aim to reconstruct the pelvis and femurs with the left and right sides separated.
Each object (hemi-bone) CT produced two depth maps (front and back faces) to train the depth model $G_d$ and segmentation model $G_s$, i.e., eight depth maps (four objects) for a single X-ray, resulting in 10604 bone objects (8626 disease-affected, 1978 healthy, as graded by \cite{masuda_automatic_2023}).
In Sec. \ref{sec:3d_no_compl}, we evaluate the 3D shapes reconstructed from estimated single- and dual- face depth maps.
Through 3D shape completion, we further show the performance difference between completion from single- and dual- face-depth-map-reconstructed 3D shapes, which we report in Sec. \ref{sec:3d_compl}.
We also compare the proposed CASI loss with conventional SI loss in depth map (2D space) in Sec. \ref{sec:depth} and reconstructed shape (3D space) in Sec. \ref{sec:3d_no_compl}.
We started from training with a low $256\times 256$ image resolution;
however, we further explore performance improvement by image resolution scaling and incorporating pre-training with Masked Autoencoder \cite{he_masked_2022} in Sec. \ref{sec:3d_compl} and Sec. \ref{sec:depth}.
A four-fold cross-validation policy was applied.
We excluded 346 (3.26\%) objects due to radiography-CT registration failure before gathering the fold results.
The segmentation model $G_{s}$ achieved a Dice score of 0.988.
To evaluate 3D shape, the average symmetric surface distance (ASSD), 95 percentile Hausdorff distance (HD95), earth mover's distance (EMD), and l2-chamfer distance ($\mathrm{CD}_{l2}$) were used.
We used mean absolute error (MAE) and root mean square error (RMSE) for depth map evaluation.

\subsection{3D shape results without shape completion}
\label{sec:3d_no_compl}
\begin{figure}
\includegraphics[width=\textwidth]{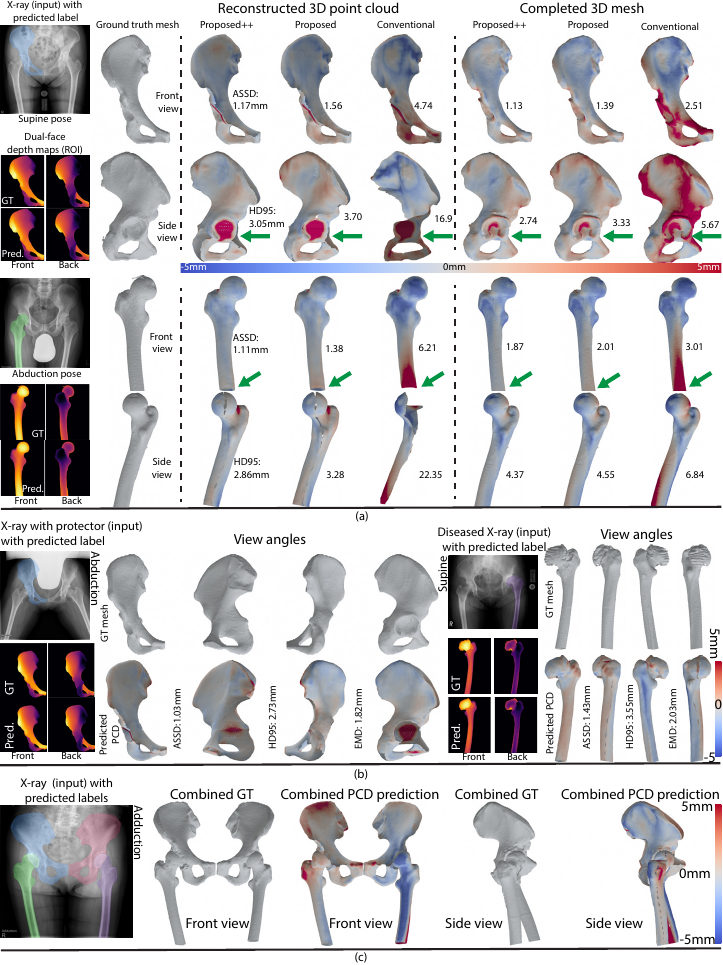}
\caption{
Visualization of the reconstructed and completed 3D shapes.
(a) Comparison between conventional (\textit{\scriptsize 256}$\Lsiindep$), the proposed (\textit{\scriptsize 256}$\Lcasiindep$), and the proposed++(\textit{\scriptsize 1024}$\Lcasiindep$\textdagger) methods.
(b) Visualization of two representative estimated samples (blocked region and slightly diseased bone) using the proposed method.
(c) A combined visualization of the target bones.
} 
\label{fig:sample_visual}
\end{figure}

Tab. \ref{tb:eval_3d} shows the evaluation results on the 3D shape reconstructed from estimated depth maps from the models trained with different settings.
The first-row method ({\scriptsize\textit{256}} $\Lsiindep$\textreferencemark) that produced single-face depth maps with $\Lsiindep$ loss is regarded as the baseline.
When predicting dual-face depth maps ({\scriptsize\textit{256}} $\Lsiindep$) with the same SI loss function significantly improved the 3D reconstruction performance, reducing the femur mean ASSD and HD95 from 5.54 and 21.1 mm to 2.28 and 5.75 mm, respectively.
This suggests that this generalization of the SI loss is useful.
The proposed CASI loss ({\scriptsize\textit{256}} $\Lcasiindep$) outperformed the conventional SI loss ({\scriptsize\textit{256}} $\Lsiindep$) on all the metrics.
We observe that the generalization without inter-depth-map pixel dependency unexpectedly performed better.
The reason for this behavior may be due to the size (thickness) difference in objects, which resulted in different error variance levels, influencing the training.
In fact, we chose not to report the results by the SI loss with pixel dependency $\Lsidep$, since the training is unstable and often fails.
The proposed CASI losses $\Lcasidep$ and $\Lcasiindep$ were always stable during training.
Fig. \ref{fig:sample_visual} shows the visual comparison between the methods on two representative samples, where the proposed methods improved the reconstruction quality significantly.

\subsection{3D shape results with shape completion}
\label{sec:3d_compl}
We use 3D completion to demonstrate the effectiveness of estimating dual-face depth.
Tab. \ref{tb:eval_3d} shows the evaluation results grouped by the disease.
The completion on the proposed dual-face method was significantly better by the fact that much richer 3D information was accessible to fitting, as shown in Fig. \ref{fig:sample_visual} (a).
The mean ASSDs were improved from 2.34 and 2.61 mm to 1.98 and 2.22 mm for the healthy and diseased pelvis, respectively.
The proposed dual-face-depth-reconstructed 3D also reduced the fitting outliers significantly indicated by CD$_{l2}$.
The mean CD$_{l2}$ values were reduced from 84.7 and 122 mm$^2$ to 19.3 and 31.2 mm$^2$ for the healthy and diseased femur, respectively.

\subsection{Depth map results}
\label{sec:depth}
To better evaluate the proposed CASI loss, we also evaluated the estimated 2D depth maps, which involved pixel-to-pixel correspondence to the ground truth depth maps. For the femur, the conventional SI loss $\Lsiindep$ and the proposed CASI loss $\Lcasiindep$ achieved a mean RMSE of 3.7 mm and 3.52 mm, respectively. For the pelvis, CASI loss improved the mean RMSE from 4.8 to 4.5 mm.
Further, bone and muscle volume estimation from an X-ray image had been studied previously by estimating 2D volume distribution \cite{gu_mskdex_2023}. 
Realizing that the 2D volume distribution is equivalent to thickness estimation at each pixel, our method with dual-face depth estimation is naturally capable of producing volume distribution by subtracting front-face depth map from back-face depth map to estimate bone volume.
Using the proposed CASI loss, the pearson correlation coefficient (PCC) between X-ray derived and CT derived pelvis volume was improved from 0.952 to 0.972 and further to 0.980 by pre-training and resolution scaling. More details can be found in supplemental materials.

\subsection{Implementation details}
The training policy was consistent across all experiments.
The AdamW optimizer \cite{loshchilov_decoupled_2018} with SGDR \cite{loshchilov_sgdr_2016} with an initial learning rate of $2\times10^{-4}$ was used, where the $T_{0}$ and $T_{i}$ were set to 10 and 2, respectively.
All the deep learning models were trained with 630 epochs, using RandAugment \cite{cubuk_randaugment_2020}.
For the depth model $G_{d}$, we used the Norm-Free Network (F0 variant) \cite{brock_high_performance_2021} as the encoder for its training high efficiency and performance.
The decoder in $G_{d}$ followed \cite{gu_bone_2023}.
We used a 2D nnU-Net \cite{isensee_nnu_net_2021} as segmentation model $G_{s}$ trained with 512 resolution.

\section{Conclusion and Summary}
In this work, we propose a new approach to the fundamentally difficult problem of 2D-3D reconstruction from a single X-ray image, termed \methodname, where we simultaneously estimate both the front an back faces of in-vivo bone structures of interest.
Furthermore, we proposed the generalization of conventional loss to multi-depth-map supervision with improvement by utilizing known geometry information.
Through rigorous experiments with large-scale X-ray dataset on real patients, we demonstrate significant improvement on 3D reconstructions. This work offers potential for many novel and established clinical applications, such as posture estimation, low X-ray dose bone disease detection, diagnosis and follow-up on widely available equipment even outside of hospitals and specialized clinics, particularly in the developing world. 
%y dual-face-depth estimation evaluated on the reconstructed 3D and 3D-completed shapes.

\begin{credits}
\subsubsection{\ackname} The research in this paper was funded by\\
MEXT/JSPS KAKENHI (19H01176, 20H04550, 21K16655).
% \subsubsection{\discintname}
% The authors have no competing interests to declare relevant to this article's content.
\end{credits}

\bibliographystyle{splncs04}
\bibliography{refs}
\end{document}

\title{Contribution Title\thanks{Supported by organization x.}}
\authorrunning{Y. Gu et al.}

\section*{Supplemental materials for \\
3DDX: Bone surface reconstruction from a single standard-geometry radiograph via dual-face depth estimation}

\begin{table}[]
\caption{
Evaluation results of depth map estimation.
\textdagger{ }denotes using pre-training.
The mean absolute error (MAE) and root mean square error (RMSE) metrics are reported in mm units.
For each metric of each bone object, we report the mean(median)$\pm$std.
}
\centering
\label{tb:depth}
\begin{tabular}{lcccc}
\hline
                & \multicolumn{2}{c|}{Femur}                 & \multicolumn{2}{c}{Pelvis}       \\
Method          & MAE $\downarrow$           & \multicolumn{1}{c|}{RMSE $\downarrow$} & MAE $\downarrow$             & RMSE $\downarrow$         \\ \hline
{\tiny\textit{256}} $\mathcal{L}_{si}^{indep}$            & 2.94(2.36)$\pm$2.11 & 3.70(3.08)$\pm$2.36            & 3.24(3.04)$\pm$0.956 & 4.80(4.53)$\pm$1.30 \\
{\tiny\textit{256}} $\mathcal{L}_{casi}^{dep}$        & 2.88(2.27)$\pm$2.29 & 3.60(2.94)$\pm$2.56            & 2.94(2.76)$\pm$0.915 & 4.53(4.26)$\pm$1.32 \\
{\tiny\textit{256}} $\mathcal{L}_{casi}^{indep}$ & 2.81(2.19)$\pm$2.28 & 3.52(2.86)$\pm$2.54            & 2.92(2.73)$\pm$0.909 & 4.50(4.23)$\pm$1.32 \\ \hline
{\tiny\textit{256}} $\mathcal{L}_{casi}^{indep}$\textdagger & 2.78(2.17)$\pm$2.27 & 3.49(2.82)$\pm$2.54            & 2.87(2.69)$\pm$0.868 & 4.47(4.20)$\pm$1.29 \\
{\tiny\textit{512}} $\mathcal{L}_{casi}^{indep}$\textdagger & 2.58(1.99)$\pm$2.14 & 3.22(2.59)$\pm$2.40            & 2.57(2.40)$\pm$0.818 & 4.05(3.77)$\pm$1.31 \\
{\tiny\textit{1024}}$\mathcal{L}_{casi}^{indep}$\textdagger & 2.54(1.95)$\pm$2.16 & 3.17(2.53)$\pm$2.43            & 2.49(2.31)$\pm$0.822 & 3.93(3.64)$\pm$1.31 \\ \hline
\end{tabular}
\end{table}

\begin{figure}
\includegraphics[width=\textwidth]{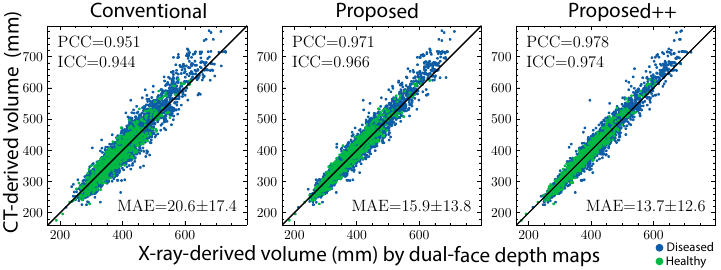}
\caption{
Scatter plot of the X-ray-derived volume against CT-derived volume.
The volume is estimated from X-ray through dual-face depth maps subtraction to obtain volume distribution map (thickness map) to calculate Volume.
The proposed method with CASI loss outperformed the conventional SI loss.
}
\label{fig:sample_visual}
\end{figure}

\begin{figure}
\includegraphics[width=\textwidth]{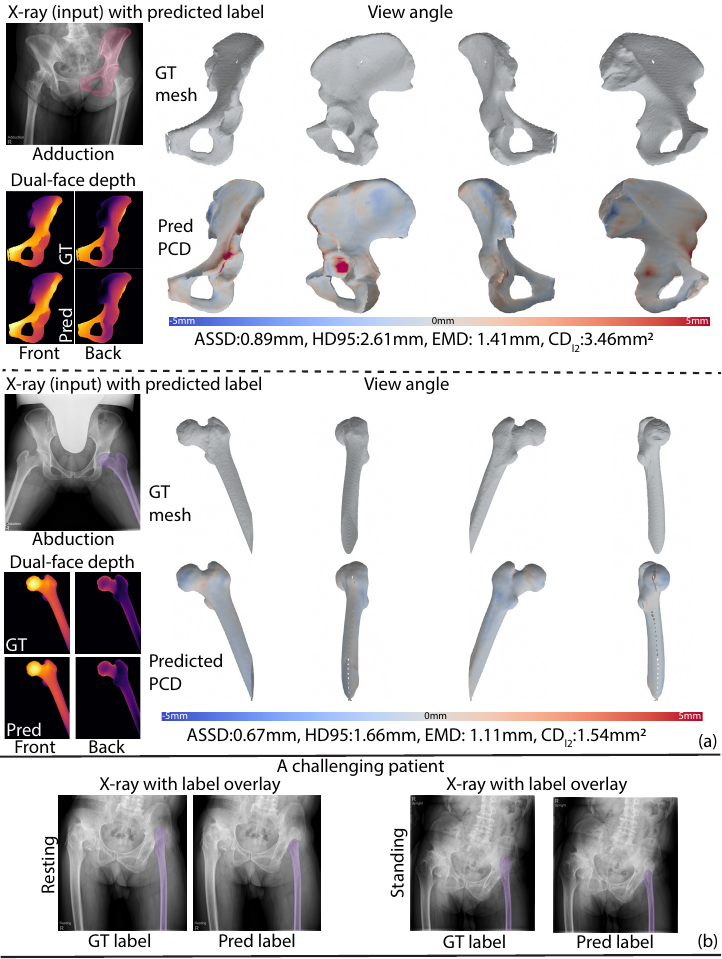}
\caption{
Visualization of two random samples by the proposed dual-face and CASI loss, using 3D completion}
\label{fig:sample_visual}
\end{figure}